\documentstyle[12pt]{article}
\newlength{\extraspace}
\newlength{\extraspaces}
\catcode`\@=11
\def\numberbysection{\@addtoreset{equation}{section}
\def\theequation{\arabic{section}.\arabic{equation}}}
\newcommand{\newsection}[1]{
\vspace{7mm}
\pagebreak[3]
\addtocounter{section}{1}
\setcounter{subsection}{0}
\setcounter{footnote}{0}
\begin{center}
{\large {\bf \thesection. #1}}
\end{center}
\nopagebreak
\medskip
\nopagebreak
\hspace{3mm}}
\newcommand{\nonu}{\nonumber \\[.5mm]}
\newcommand{\A}{&\!\!\!}

\newcommand{\be}{\begin{equation}}
\newcommand{\bea}{\begin{eqnarray}}
\newcommand{\eea}{\end{eqnarray}}
\newcommand{\ba}{\begin{array}}
\newcommand{\ea}{\end{array}}
\newcommand{\ee}{\end{equation}}

\newcommand{\br}{\begin{array}}
\newcommand{\er}{\end{array}}
\begin{document}
\addtolength{\baselineskip}{.7mm}
\thispagestyle{empty}
\begin{flushright}
MIT--CTP--3324 \\
STUPP--02--172 \\
{\tt hep-th/0212337} \\ 
December, 2002
\end{flushright}
\vspace{7mm}
\begin{center}
{\Large{\bf Three-Form Flux with ${\cal N}=2$ Supersymmetry \\[2mm]
on AdS${}_5$ $\times$ S${}^5$ 
}} \\[20mm] 
{\sc Madoka Nishimura}${}^{\rm a}$\footnote{
\tt e-mail: madoka@mit.edu} 
\hspace{1mm} and \hspace{2mm}
{\sc Yoshiaki Tanii}${}^{\rm b}$\footnote{
\tt e-mail: tanii@post.saitama-u.ac.jp} \\[7mm]
${}^{\rm a}${\it Center for Theoretical Physics \\
Massachusetts Institute of Technology \\
Cambridge, MA 02139-4307, USA} \\[3mm]
${}^{\rm b}${\it Physics Department, Faculty of Science \\
Saitama University, Saitama 338-8570, Japan} \\[20mm]
{\bf Abstract}\\[7mm]
{\parbox{13cm}{\hspace{5mm}
In the context of the AdS/CFT correspondence the general form of 
a three-form flux perturbation to the AdS${}_5$ $\times$ S${}^5$ 
solution in the type IIB supergravity which preserves ${\cal N}=2$ 
supersymmetry is obtained. 
The arbitrary holomorphic function appearing in the ${\cal N}=1$ 
case studied by Gra\~na and Polchinski is restricted to a quadratic 
function of the coordinates transverse to the D3-branes. 
}}
\end{center}
\vfill
\newpage
\setcounter{section}{0}
\setcounter{equation}{0}
%
\newsection{Introduction}
%
It was proposed that the type IIB string theory compactified on 
AdS${}_5$ $\times$ S${}^5$ has a dual description by the 
${\cal N}=4$ super Yang-Mills theory in the large $N$ limit 
\cite{MAL,GKP,WITTEN}. 
This conjecture of the AdS/CFT correspondence has been supported 
by comparison of spectra, correlation functions and anomalies 
calculated in both of the supergravity and the Yang-Mills 
theory. (For a review, see ref.\ \cite{AGMOO}.) 
The AdS/CFT correspondence was also studied in various other 
spacetime dimensions. 
At first the correspondence was studied for theories with high 
supersymmetries such as ${\cal N}=4$. To apply it to more realistic 
models one has to consider theories with lower supersymmtries. 
\par
One of the ways to obtain the AdS/CFT correspondence for lower 
supersymmetric cases is to modify supergravity solutions by adding 
a perturbation. In ref.\ \cite{PS} a perturbation of the three-form 
flux was added to the AdS${}_5$ $\times$ S${}^5$, which breaks 
${\cal N}=4$ to ${\cal N}=1$. This perturbation corresponds to 
fermion mass terms of the three ${\cal N}=1$ chiral multiplets 
in the ${\cal N}=4$ super Yang-Mills theory and polarizes 
D3 branes into 5-branes \cite{MYERS,TVR}. 
Similar constructions of the AdS/CFT correspondence with lower 
supersymmetries were discussed in refs.\ \cite{BENA,BN1,BN2,NISHI}. 
\par
The general form of a three-form flux perturbation to the 
AdS${}_5$ $\times$ S${}^5$ solution which preserves ${\cal N}=1$ 
supersymmetry and satisfies the Bianchi identity and the linearized 
field equation was obtained in ref.\ \cite{GP}. It contains an 
arbitrary holomorphic function and an arbitrary harmonic function 
of the coordinates for the directions transverse to the D3-branes. 
It was argued that the holomorphic function corresponds to a 
superpotential in the dual super Yang-Mills theory. 
When the holomorphic function is quadratic in the transverse 
coordinates, the three-form flux coincides with that of 
ref.\ \cite{PS}. 
\par
The purpose of the present paper is to obtain the general form of 
a three-form flux perturbation to the AdS${}_5$ $\times$ S${}^5$ 
solution which preserves ${\cal N}=2$ supersymmetry. 
We use the result of the ${\cal N}=1$ case \cite{GP} and require 
further that the second supersymmetry is preserved. 
We find that the arbitrary holomorphic function in the 
${\cal N}=1$ case is restricted to a quadratic function of the 
transverse coordinates. This is a special form of the perturbation 
studied in ref.\ \cite{PS}, which has one vanishing mass. 
It would be interesting to study a relation of our result to 
other works on soft breaking of ${\cal N}=4$ to ${\cal N}=2$ 
in the Coulomb branch \cite{PW,BPP,BEH}. 
In order to discuss the corresponding dual field theory and 
its RG flows we need to find out an exact solution with 
non-vanishing three-form flux. 
In addition, it would be also interesting to discuss the brane 
representations and massive vacua using S-dual transformations. 
\par
%
\newsection{Unperturbed solution}
%
The field content of the type IIB supergravity in ten 
dimensions \cite{SCHWARZ,HW} is a metric $g_{MN}$, 
a complex Rarita-Schwinger field $\psi_M$, 
a real fourth-rank antisymmetric tensor field with a self-dual 
field strength $F_{MNPQR}$, a complex second-rank antisymmetric 
tensor field with a field strength $G_{MNP}$, 
a complex spinor field $\lambda$ and a complex scalar filed 
$\tau = C + i e^{-\Phi}$. We denote ten-dimensional world indices 
as $M,N,\cdots = 0,1,\cdots,9$ and local Lorentz indices as 
$A,B, \cdots = 0,1,\cdots, 9$. 
The fermionic fields satisfy chirality conditions 
$\bar\Gamma_{10D} \psi_M = \psi_M$, 
$\bar\Gamma_{10D} \lambda = - \lambda$, 
where $\bar\Gamma_{10D} = \Gamma^0 \Gamma^1 \cdots \Gamma^9$ 
is the ten-dimensional chirality matrix. 
We choose the ten-dimensional gamma matrices $\Gamma^A$ to 
have real components. 
\par
The field equations of this theory have a solution with the 
AdS${}_5$ $\times$ S${}^5$ metric \cite{GM,KRvN} 
\begin{equation}
g_{MN} dx^M dx^N 
= Z^{-{1 \over 2}} \eta_{\mu\nu} dx^\mu dx^\nu 
+ Z^{1 \over 2} \delta_{mn} dx^m dx^n, 
\label{metric}
\end{equation}
where $M=(\mu,m)$ ($\mu=0,1,2,3$; $m=4,5,\cdots,9$), 
$Z = {R^4 \over r^4}$ and $r^2 = x^m x^n \delta_{mn}$. 
The constant $R$ is a radius of AdS${}_5$ and S${}^5$. 
The fifth-rank field strength has non-vanishing components 
\begin{eqnarray}
F_{\mu\nu\rho\sigma m} 
\A = \A {1 \over \kappa Z^2} \epsilon_{\mu\nu\rho\sigma} 
\partial_m Z, \nonu
F_{mnpqr} 
\A = \A - {Z^{1 \over 2} \over \kappa} \epsilon_{mnpqrs} \partial^s Z,  
\label{fiftht}
\end{eqnarray}
where $\kappa$ is a coupling constant. 
This solution represents a supergravity configuration produced 
by D3-branes located at $x^m = 0$. 
More generally, the warp factor $Z$ can be an arbitrary function 
of $x^m$ which is harmonic except at points where D3-branes exist. 
We will consider the general $Z$ but assume that the density of 
D3-branes vanishes for $r \rightarrow \infty$ and therefore 
$Z \rightarrow {R^4 \over r^4}$ for $r \rightarrow \infty$. 
\par
We are interested in how many supersymmetries are preserved 
by this solution and by a solution with a perturbation of $G_{MNP}$ 
discussed later. They are found by studying vanishing of local 
supertransformations of the fermionic fields $\psi_M$ and $\lambda$. 
The supertransformations of the fermionic fields \cite{SCHWARZ,HW} 
in these backgrounds are 
\begin{eqnarray}
\delta \psi_M 
\A = \A {1 \over \kappa} D_M \epsilon 
+ {1 \over 16 \cdot 5!} i F_{P_1 \cdots P_5} 
\Gamma^{P_1 \cdots P_5} \Gamma_M \epsilon 
- {1 \over 96} G_{NPQ} \left( \Gamma_M{}^{NPQ} 
- 9 \delta_M^N \Gamma^{PQ} \right) \epsilon^*, \nonu
\delta \lambda 
\A = \A {1 \over 24} G_{MNP} \Gamma^{MNP} \epsilon,
\label{susytrans}
\end{eqnarray}
where the transformation parameter $\epsilon$ is a complex spinor 
satisfying the chirality condition 
$\bar\Gamma_{10D} \epsilon = \epsilon$. 
To study the supertransformations for the above backgrounds 
it is convenient to represent the ten-dimensional gamma matrices as 
\begin{eqnarray}
\Gamma^\mu \A = \A \gamma^\mu \otimes {\bf 1}, \nonu
\Gamma^m \A = \A \bar\gamma_{4D} \otimes \gamma^m, 
\end{eqnarray}
where $\gamma^\mu$ and $\gamma^m$ are the SO(3,1) and SO(6) gamma 
matrices respectively. The chirality matrices are defined as 
\begin{equation}
\bar\gamma_{4D} = i \gamma^0 \gamma^1 \gamma^2 \gamma^3, \qquad
\bar\gamma_{6D} = i \gamma^4 \gamma^5 \gamma^6 \gamma^7 
\gamma^8 \gamma^9, 
\end{equation}
which are related to the ten-dimensional one as 
$\bar\Gamma_{10D} = - \bar\gamma_{4D} \bar\gamma_{6D}$. 
\par
The above solution (\ref{metric}), (\ref{fiftht}) without 
a perturbation has 32 supersymmetries \cite{GM,KRvN}. 
This can be seen as follows. The supertransformation 
$\delta\lambda$ automatically vanishes, 
while the vanishing of $\delta\psi_M$ requires 
\begin{equation}
\tilde{D}_M \epsilon = 0, 
\label{killing}
\end{equation}
where we have defined 
\begin{eqnarray}
\tilde{D}_\mu \A = \A \partial_\mu - {1 \over 8Z} \partial_m Z 
\gamma_\mu \gamma^m (1+\bar\gamma_{4D}), \nonu
\tilde{D}_m \A = \A \partial_m - {1 \over 8Z} \partial_n Z \left( 
\delta_m^n \bar\gamma_{4D} - \gamma_m{}^n (1+\bar\gamma_{4D}) \right). 
\label{tilded}
\end{eqnarray}
For solutions of eq.\ (\ref{killing}) to exist the integrability 
condition 
\begin{equation}
[ \tilde{D}_M, \tilde{D}_N ] \epsilon = 0
\label{intcond}
\end{equation}
must be satisfied. Using the expression (\ref{tilded}) it is easy to 
show that eq.\ (\ref{intcond}) is satisfied for an arbitrary $\epsilon$. 
Therefore, all of 32 supersymmetries are preserved \cite{GM,KRvN}. 
{}From the four-dimensional field theoretical point of view 
in the AdS/CFT correspondence 
16 of them are Poincar\'e supersymmetries 
while other 16 are conformal supersymmetries. 
Thus, we have ${\cal N}=4$ supersymmetry in four dimensions. 
More explicitly, the solutions of eq.\ (\ref{killing}) with the 
chirality $\bar\gamma_{4D} = -1$ have a form 
\begin{equation}
\epsilon = Z^{-{1 \over 8}} \eta, 
\end{equation}
where $\eta$ is an arbitrary constant spinor with the chirality 
$\bar\gamma_{4D} = -1$. 
These solutions correspond to Poincar\'e supersymmetries. 
The solutions with the chirality $\bar\gamma_{4D} = +1$ depend 
on $x^\mu$ and correspond to conformal supersymmetries. 
\par
%
\newsection{Three-form flux with ${\cal N}=2$ supersymmetry}
%
By introducing a perturbation of the three-form flux $G_{mnp}$ 
the ${\cal N}=4$ supersymmetry of the unperturbed supergravity 
background is broken to lower ${\cal N}$. 
In ref.\ \cite{GP} the conditions on $G_{mnp}$ 
for unbroken ${\cal N}=1$ supersymmetry were studied. 
The supersymmetry parameter is expanded as 
$\epsilon = \epsilon_0 + \epsilon_1 + \cdots$, where $\epsilon_0$ 
is a solution of eq.\ (\ref{killing}) for the unperturbed background 
and $\epsilon_1$ is the first order correction due to the perturbation. 
Substituting it into eq.\ (\ref{killing}) $\epsilon_1$ is determined 
by $\epsilon_0$. 
To proceed it is convenient to define complex coordinates 
$z^i$ ($i=1,2,3$) from $x^m$ 
\begin{equation}
z^1 = {1 \over \sqrt{2}} (x^4+ix^7), \quad
z^2 = {1 \over \sqrt{2}} (x^5+ix^8), \quad
z^3 = {1 \over \sqrt{2}} (x^6+ix^9). 
\end{equation}
\par
It was required in ref.\ \cite{GP} that one of the four Poincar\'e 
supersymmetries $\epsilon_0 = Z^{-{1 \over 8}} \eta$, where $\eta$ 
is a constant spinor satisfying 
\begin{equation}
\gamma^{\bar{1}} \eta = \gamma^{\bar{2}} \eta 
= \gamma^{\bar{3}} \eta = 0,  
\label{gammacond1}
\end{equation}
is preserved. Here, $\bar{i}$ denote indices of $\bar{z}^i$, 
while $i$ denote those of $z^i$. 
Using the expression $\bar\gamma_{6D} = (1-\gamma^1\gamma^{\bar{1}}) 
(1-\gamma^2\gamma^{\bar{2}}) (1-\gamma^3\gamma^{\bar{3}})$ it is easy 
to see that this $\epsilon_0$ has the chirality $\bar\gamma_{4D} 
= -\bar\gamma_{6D} = -1$ appropriate for the Poincar\'e supersymmetry. 
Then, this ${\cal N}=1$ supersymmetry restricts the form of $G_{mnp}$ 
as \cite{GP}
\begin{eqnarray}
G_{ijk} \A = \A 0, \nonu
G_{ij\bar{k}} \A = \A {2 \over 3} \hat\epsilon_{\bar{k}}{}^{pq} 
\partial^{-2} \partial_p \partial_{[i} \phi 
\partial_{j]} \partial_q Z 
+ \hat\epsilon_{ij}{}^{\bar{l}} \partial_{\bar{k}} 
\partial_{\bar{l}} \psi, \nonu
G_{i\bar{j}\bar{k}} \A = \A {1 \over 12} 
\hat\epsilon_{\bar{j}\bar{k}}{}^l \left( 
2 \partial_i \partial_l \phi Z 
- \alpha \hat\epsilon_{il}{}^{\bar{k}} \partial_{\bar{k}} Z 
- 4 \partial_{[i} \phi \partial_{l]} Z \right), \nonu
G_{\bar{i}\bar{j}\bar{k}} \A = \A {1 \over 6} 
\hat\epsilon_{\bar{i}\bar{j}\bar{k}} \delta^{l\bar{l}} \partial_l \phi 
\partial_{\bar{l}} Z, 
\label{gp}
\end{eqnarray}
where $\phi(z^1, z^2, z^3)$ is an arbitrary holomorphic function, 
$\alpha$ is an arbitrary constant and $\psi$ is an arbitrary harmonic 
function.\footnote{In ref.\ \cite{GP} the constant $\alpha$ is required 
to vanish by the Bianchi identity. However, we do not agree with this 
result and leave $\alpha$ non-vanishing.} 
In eq.\ (\ref{gp}) $\hat\epsilon_{ij}{}^{\bar{k}}$ and 
$\hat\epsilon_{\bar{i}\bar{j}}{}^k$ are totally antisymmetric in their 
indices and take constant values $0$, $\pm 1$ regardless of index 
positions, and $\partial^2 
= 2 \delta^{i\bar{i}} \partial_i \partial_{\bar{i}}$ is the Laplacian. 
The three-form flux (\ref{gp}) also satisfies the Bianchi identity as 
well as the linearized field equation. 
\par
We shall obtain conditions on $G_{mnp}$ for unbroken ${\cal N}=2$ 
supersymmetry. We require that 
in addition to $\epsilon_0 = Z^{-{1 \over 8}} \eta$ the second 
supersymmetry with the parameter 
\begin{equation}
\epsilon_0 = Z^{-{1 \over 8}} \gamma^1 \gamma^2 \eta
\end{equation}
is also preserved. This $\epsilon_0$ satisfies 
\begin{equation}
\gamma^1 \epsilon_0 = \gamma^2 \epsilon_0 
= \gamma^{\bar{3}} \epsilon_0 = 0 
\label{gammacond2}
\end{equation}
and has the chirality $\bar\gamma_{4D} = -1$. 
Comparing eqs.\ (\ref{gammacond1}) and (\ref{gammacond2}) it is easy 
to see that the conditions for the second supersymmetry 
are obtained from eq.\ (\ref{gp}) by the replacements 
\begin{equation}
1 \leftrightarrow \bar{1}, \quad
2 \leftrightarrow \bar{2}, \quad
\alpha \rightarrow \alpha', \quad
\phi(z^1, z^2, z^3) 
\rightarrow \phi'(\bar{z}^1, \bar{z}^2, z^3), \quad
\psi \rightarrow \psi'
\label{replace}
\end{equation}
for new $\alpha'$, $\phi'$ and $\psi'$. 
\par
We now require that the expression (\ref{gp}) and that with 
the replacements (\ref{replace}) are compatible each other. 
Let us first consider $G_{123}$. From the expression (\ref{gp}) 
we have $G_{123} = 0$. From the other expression we have 
$G_{123} = {1 \over 6} \partial_3^2 \phi' Z$, which is obtained from 
$G_{\bar{1}\bar{2}3}$ in eq.\ (\ref{gp}) by the replacements 
(\ref{replace}). Thus we obtain a condition 
\begin{equation}
G_{123} : \quad \partial_3^2 \phi' = 0. 
\end{equation}
Similarly, we obtain conditions 
\begin{eqnarray}
G_{2\bar{2}1} + G_{3\bar{3}1} \A : \A \quad 
\partial_{\bar{2}} \partial_3 \phi' = 0, \nonu
G_{1\bar{1}2} + G_{3\bar{3}2} \A : \A \quad 
\partial_{\bar{1}} \partial_3 \phi' = 0, \nonu
G_{\bar{1}\bar{2}3} \A : \A \quad 
\partial_3^2 \phi = 0, \nonu
G_{\bar{2}2\bar{1}} + G_{3\bar{3}\bar{1}} \A : \A \quad
\partial_2 \partial_3 \phi = 0, \nonu
G_{\bar{1}1\bar{2}} + G_{3\bar{3}\bar{2}} \A : \A \quad
\partial_1 \partial_3 \phi = 0, \nonu
G_{1\bar{2}\bar{3}} \A : \A \quad
\partial_1^2 \phi = \partial_{\bar{2}}^2 \phi', \nonu
G_{\bar{1}2\bar{3}} \A : \A \quad
\partial_2^2 \phi = \partial_{\bar{1}}^2 \phi'. 
\label{cond1}
\end{eqnarray}
The component $G_{1\bar{1}3} + G_{2\bar{2}3}$ vanishes in both of the 
two expressions and gives no condition. 
These conditions fix the forms of $\phi$ and $\phi'$ as 
\begin{eqnarray}
\phi \A = \A m_1 (z^1)^2 + m_2 (z^2)^2 + 2a z^1 z^2 
+ b_1 z^1 + b_2 z^2 + b_3 z^3, \nonu
\phi' \A = \A m_2 (\bar{z}^1)^2 + m_1 (\bar{z}^2)^2 
+ 2a' \bar{z}^1 \bar{z}^2 + b'_1 \bar{z}^1 + b'_2 \bar{z}^2 
+ b'_3 z^3, 
\label{bilinear}
\end{eqnarray}
where $m_1$, $m_2$, $a$, $a'$, $b_i$ and $b'_i$ are arbitrary 
constants. We further obtain conditions 
\begin{eqnarray}
G_{\bar{1}23} \A : \A \quad
\partial_{\bar{1}}^2 \psi = \partial_2^2 \psi', \nonu
G_{1\bar{2}3} \A : \A \quad
\partial_{\bar{2}}^2 \psi = \partial_1^2 \psi', \nonu
G_{31\bar{1}} \A : \A \quad
\partial_{\bar{1}} \partial_{\bar{2}} \psi 
= - \partial_1 \partial_2 \psi', \qquad a = -a'. 
\label{cond4}
\end{eqnarray}
By a linear transformation $z^i \rightarrow U^i{}_j z^j$ 
($i,j = 1,2$) with a unitary matrix $U$ we can set $a = -a' = 0$. 
\par
So far we have not used a particular form of $Z$. 
We now examine the remaining conditions first by using 
the asymptotic form $Z \sim {R^4 \over r^4}$ for 
$r \rightarrow \infty$ to fix the coefficients in 
eq.\ (\ref{bilinear}) and $\alpha$, $\alpha'$. We then check 
that the conditions are satisfied also for $r < \infty$. 
{}From the equation for $G_{1\bar{1}\bar{3}}$ we obtain 
\begin{eqnarray}
G_{1\bar{1}\bar{3}} : \;
\A\A - {1 \over 6} \partial_1 \partial_2 \phi Z 
+ {1 \over 12} \left( \alpha \partial_{\bar{3}} Z 
+ 2 \partial_1 \phi \partial_2 Z 
-  2 \partial_2 \phi \partial_1 Z \right) \nonu
\A\A \qquad\qquad = 
{1 \over 6} \partial_{\bar{1}} \partial_{\bar{2}} \phi' Z 
- {1 \over 12} \left( \alpha' \partial_{\bar{3}} Z 
+ 2 \partial_{\bar{1}} \phi' \partial_{\bar{2}} Z 
-  2 \partial_{\bar{2}} \phi' \partial_{\bar{1}} Z \right). 
\label{cond2}
\end{eqnarray}
The equation for $G_{2\bar{2}\bar{3}}$ gives the same condition. 
Substituting the asymptotic form $Z \sim {R^4 \over r^4}$ and 
eq.\ (\ref{bilinear}) into eq.\ (\ref{cond2}) we find 
$\alpha' = - \alpha$ and $b_1 = b_2 = b'_1 = b'_2 = 0$. 
The remaining conditions become 
\begin{eqnarray}
G_{1\bar{1}\bar{2}} \A : \A \quad
\partial_1 \partial_{\bar{3}} \psi' 
= {1 \over 12} ( \alpha \partial_{\bar{2}} 
+ 2 b_3 \partial_1 ) Z, \nonu
G_{3\bar{3}\bar{1}} \A : \A \quad
\partial_2 \partial_{\bar{3}} \psi' 
= - {1 \over 12} ( \alpha \partial_{\bar{1}} 
- 2 b_3 \partial_2 ) Z, \nonu
G_{\bar{1}\bar{2}\bar{3}} \A : \A \quad
\partial_{\bar{3}}^2 \psi' 
= {1 \over 6} b_3 \partial_{\bar{3}} Z, \nonu
G_{23\bar{3}} \A : \A \quad
\partial_{\bar{1}} \partial_{\bar{3}} \psi 
= - {1 \over 12} ( \alpha \partial_2 
- 2 b'_3 \partial_{\bar{1}} ) Z, \nonu
G_{12\bar{2}} \A : \A \quad
\partial_{\bar{2}} \partial_{\bar{3}} \psi 
= {1 \over 12} ( \alpha \partial_1 
+ 2 b'_3 \partial_{\bar{2}} ) Z, \nonu
G_{12\bar{3}} \A : \A \quad
\partial_{\bar{3}}^2 \psi 
= {1 \over 6} b'_3 \partial_{\bar{3}} Z. 
\label{cond3}
\end{eqnarray}
Comparing the equation obtained by applying $\partial_{\bar{3}}$ 
to the first equation in eq.\ (\ref{cond3}) and that obtained 
by applying $\partial_1$ to the third equation we find $\alpha = 0$. 
Then, eq.\ (\ref{cond3}) determines $\psi$, $\psi'$ as 
\begin{eqnarray}
\partial_{\bar{3}} \psi 
\A = \A {1 \over 6} b'_3 Z + f(z^1,z^2,z^3), \nonu
\partial_{\bar{3}} \psi' 
\A = \A {1 \over 6} b_3 Z + f'(\bar{z}^1,\bar{z}^2,z^3), 
\label{psis}
\end{eqnarray}
where $f$ and $f'$ are arbitrary functions of each variables. 
Substituting eq.\ (\ref{psis}) into the $\bar{z}^3$ derivative of 
eq.\ (\ref{cond4}) and using the asymptotic form 
$Z \sim {R^4 \over r^4}$ we obtain $b_3 = b'_3 = 0$. 
\par
{}As a result of these analyses at asymptotic region $r \sim \infty$ 
we obtain 
\begin{eqnarray}
\phi \A = \A m_1 (z^1)^2 + m_2 (z^2)^2, \nonu
\phi' \A = \A m_2 (\bar{z}^1)^2 + m_1 (\bar{z}^2)^2. 
\label{bilinear2}
\end{eqnarray}
We have to check that eqs.\ (\ref{cond4}), (\ref{cond2}) and 
(\ref{cond3}) are satisfied even for $r < \infty$. 
Substituting eq.\ (\ref{bilinear2}) into eq.\ (\ref{cond3})
we find that their right-hand sides vanish. 
The general solution of these equations are 
\begin{eqnarray}
\psi \A = \A f(z^1,z^2,z^3) \bar{z}^3 
+ g(z^1,\bar{z}^1,z^2,\bar{z}^2,z^3), \nonu
\psi' \A = \A f'(\bar{z}^1,\bar{z}^2,z^3) \bar{z}^3 
+ g'(z^1,\bar{z}^1,z^2,\bar{z}^2,z^3), 
\label{psis2}
\end{eqnarray}
where $f$, $f'$, $g$ and $g'$ are arbitrary functions of each 
variables. The conditions in eq.\ (\ref{cond4}) then require 
\begin{equation}
\partial_{\bar{1}}^2 g = \partial_2^2 g', \qquad
\partial_{\bar{2}}^2 g = \partial_1^2 g', \qquad
\partial_{\bar{1}} \partial_{\bar{2}} g = - \partial_1 \partial_2 g'. 
\label{fgcond}
\end{equation}
The conditions that $\psi$ and $\psi'$ in eq.\ (\ref{psis2}) 
are harmonic are 
\begin{eqnarray}
\partial^2 g(z^1,\bar{z}^1,z^2,\bar{z}^2,z^3)
\A = \A - \partial_3 f(z^1,z^2,z^3), \nonu
\partial^2 g'(z^1,\bar{z}^1,z^2,\bar{z}^2,z^3)
\A = \A - \partial_3 f'(\bar{z}^1,\bar{z}^2,z^3). 
\label{fgcond2}
\end{eqnarray}
The functions $f$ and $f'$ do not appear in $G_{mnp}$ 
as one can see by substituting eq.\ (\ref{psis2}) into eq.\ (\ref{gp}). 
We only need to consider $g$ and $g'$. 
Eq.\ (\ref{fgcond2}) means that $\partial^2 g$ and 
$\partial^2 g'$ are independent of $\bar{z}^1, \bar{z}^2$ and 
$z^1, z^2$ respectively. These conditions are automatically 
satisfied when $g$ and $g'$ satisfy eq.\ (\ref{fgcond}).
The functions $g$ and $g'$ need not be harmonic. 
Finally, we have to consider eq.\ (\ref{cond2}). 
Substituting eq.\ (\ref{bilinear2}) into eq.\ (\ref{cond2}) 
we obtain 
\begin{equation}
\left( m_1 z^1 \partial_2 - m_2 z^2 \partial_1 
+ m_2 \bar{z}^1 \partial_{\bar{2}}
- m_1 \bar{z}^2 \partial_{\bar{1}} \right) Z = 0. 
\label{rotinv}
\end{equation}
This means that $Z$ is invariant under SO(2) rotation of 
$(\sqrt{m_1} \, z^1, \sqrt{m_2} \, z^2)$ and 
$(\sqrt{m_2} \, \bar{z}^1, \sqrt{m_1} \, \bar{z}^2 )$. 
Therefore, $Z$ must be a function of SO(2) invariant variables 
$r^2 = 2 \left( z^1 \bar{z}^1 + z^2 \bar{z}^2 \right)$, 
$m_1 (z^1)^2 + m_2 (z^2)^2$, 
$m_2 (\bar{z}^1)^2 + m_1 (\bar{z}^2)^2$ and 
$m_1 z^1 \bar{z}^2 - m_2 z^2 \bar{z}^1$. 
\par
Let us summarize the result. 
The general form of the three-form flux $G_{mnp}$ which preserves 
the ${\cal N}=2$ supersymmetry at the first order of the perturbation 
is given by eq.\ (\ref{gp}) with $\alpha = 0$, $\phi$ 
in eq.\ (\ref{bilinear2}) and $\psi$ replaced by 
$g(z^1,\bar{z}^1,z^2,\bar{z}^2,z^3)$ satisfying eq.\ (\ref{fgcond}) 
for some function $g'(z^1,\bar{z}^1,z^2,\bar{z}^2,z^3)$. 
Thus, $\phi$, which is an arbitrary holomorphic function in the 
${\cal N}=1$ case \cite{GP}, is severely restricted to a quadratic 
function in the ${\cal N}=2$ case. 
Such ${\cal N}=2$ preserving perturbation is possible only when 
the warp factor $Z$ satisfies eq.\ (\ref{rotinv}). 
\par
In our analysis at the first order of the perturbation we did not 
need the condition $m_1 = m_2$ to obtain the ${\cal N}=2$ 
supersymmetry. At higher orders \cite{FM} we would need the 
condition $m_1 = m_2$ since these parameters correspond to masses 
of two ${\cal N}=1$ chiral multiplets, which should be combined 
into an ${\cal N}=2$ hypermultiplet. 
This is indeed the case in the field theory side. 
To see this let us consider two ${\cal N}=1$ chiral supermultiplets 
$(A_1, \psi_1)$ and $(A_2, \psi_2)$, where $A_1$, $A_2$ are complex 
scalar fields and $\psi_1$, $\psi_2$ are Weyl spinor fields, 
with the action 
\begin{eqnarray}
S \A = \A \int d^4 x \biggl[ - \partial_\mu A_1^* \partial^\mu A_1 
- \partial_\mu A_2^* \partial^\mu A_2 
- i \psi_1 \sigma^\mu \partial_\mu \bar\psi_1 
- i \psi_2 \sigma^\mu \partial_\mu \bar\psi_2 \nonu
\A\A 
- m_1^2 A_1^* A_1 
- m_2^2 A_2^* A_2 
- {1 \over 2} m_1 \left( \psi_1 \psi_1 
+ \bar{\psi}_1 \bar{\psi}_1 \right)
- {1 \over 2} m_2 \left( \psi_2 \psi_2 
+ \bar{\psi}_2 \bar{\psi}_2 \right) \biggr]. 
\end{eqnarray}
Here we have used the two-component spinor notation in ref.\ \cite{WB}. 
$S$ is invariant under the ${\cal N}=1$ supertransformation 
\begin{equation}
\delta A_i = \sqrt{2} \epsilon \psi_i, \qquad
\delta \psi_i = \sqrt{2} i \sigma^\mu \bar\epsilon \partial_\mu A_i 
- \sqrt{2} m_i \epsilon A_i^* \qquad (i=1,2). 
\end{equation}
The exact $N=2$ supersymmetry of course requires $m_1 = m_2$. 
However, even for $m_1 \not= m_2$, it is also invariant under the 
second supertransformation 
\begin{eqnarray}
\delta A_1 \A = \A \sqrt{2} \epsilon \psi_2, \qquad
\delta \psi_1 = \sqrt{2} i \sigma^\mu \bar\epsilon \partial_\mu A_2 
- \sqrt{2} m_1 \epsilon A_2^*, \nonu
\delta A_2 \A = \A - \sqrt{2} \epsilon \psi_1, \qquad
\delta \psi_2 = - \sqrt{2} i \sigma^\mu \bar\epsilon \partial_\mu A_1 
+ \sqrt{2} m_2 \epsilon A_1^* 
\end{eqnarray}
at the first order in $m_1$, $m_2$. 
Thus, the condition $m_1 = m_2$ is needed only in quadratic and 
higher order terms for the ${\cal N}=2$ supersymmetry. 
\par
\newpage
\vspace{10mm}
\noindent {\Large{\bf Acknowledgements}} 
\vspace{3mm}

One of the authors (M.N.) would like to thank N. Constable, 
D.Z. Freedman, A. Hanany, W. Taylor, J. Troost and A. Tsuchiya 
for helpful discussions, and the Center for Theoretical Physics 
of MIT for hospitality. 
This work is supported in part by the Nishina Memorial 
Foundation and by funds provided by the U.S. Department of Energy 
(D.O.E.) under cooperative research agreement \#DF-FC02-94ER40818. 
%
%
\newcommand{\NP}[1]{{\it Nucl.\ Phys.\ }{\bf #1}}
\newcommand{\PL}[1]{{\it Phys.\ Lett.\ }{\bf #1}}
\newcommand{\CMP}[1]{{\it Commun.\ Math.\ Phys.\ }{\bf #1}}
\newcommand{\MPL}[1]{{\it Mod.\ Phys.\ Lett.\ }{\bf #1}}
\newcommand{\IJMP}[1]{{\it Int.\ J. Mod.\ Phys.\ }{\bf #1}}
\newcommand{\PR}[1]{{\it Phys.\ Rev.\ }{\bf #1}}
\newcommand{\PRL}[1]{{\it Phys.\ Rev.\ Lett.\ }{\bf #1}}
\newcommand{\PTP}[1]{{\it Prog.\ Theor.\ Phys.\ }{\bf #1}}
\newcommand{\PTPS}[1]{{\it Prog.\ Theor.\ Phys.\ Suppl.\ }{\bf #1}}
\newcommand{\AP}[1]{{\it Ann.\ Phys.\ }{\bf #1}}
\newcommand{\ATMP}[1]{{\it Adv.\ Theor.\ Math.\ Phys.\ }{\bf #1}}
\end{document}